\documentclass[%
 reprint,
superscriptaddress,
 amsmath,amssymb,
 aps,
 pra,
]{revtex4-2}

\usepackage{graphicx}
\usepackage{dcolumn}
\usepackage{bm}
\usepackage{xcolor}

\newcommand{\grtHz}{$g/\sqrt{\mathrm{Hz}}$}

\begin{document}

\preprint{APS/123-QED}

\title{Precision optomechanical accelerometer via hybrid test mass integration}

\author{Nathaniel Bawden}
\affiliation{School of Mathematics and Physics, The University of Queensland, St Lucia, Queensland 4067, Australia}
\affiliation{ARC Centre of Excellence for Engineered Quantum Systems, St Lucia, Queensland 4067, Australia}
\affiliation{ARC Centre of Excellence in Quantum Biotechnology, St. Lucia, Queensland 4067, Australia}

\author{Benjamin J. Carey}
\affiliation{School of Mathematics and Physics, The University of Queensland, St Lucia, Queensland 4067, Australia}
\affiliation{ARC Centre of Excellence for Engineered Quantum Systems, St Lucia, Queensland 4067, Australia}
\affiliation{ARC Centre of Excellence in Quantum Biotechnology, St. Lucia, Queensland 4067, Australia}

\author{Poh-Meng Yeo}
\affiliation{School of Mathematics and Physics, The University of Queensland, St Lucia, Queensland 4067, Australia}
\affiliation{ARC Centre of Excellence for Engineered Quantum Systems, St Lucia, Queensland 4067, Australia}
\affiliation{ARC Centre of Excellence in Quantum Biotechnology, St. Lucia, Queensland 4067, Australia}

\author{Nishta Arora}
\affiliation{School of Mathematics and Physics, The University of Queensland, St Lucia, Queensland 4067, Australia}
\affiliation{ARC Centre of Excellence for Engineered Quantum Systems, St Lucia, Queensland 4067, Australia}
\affiliation{ARC Centre of Excellence in Quantum Biotechnology, St. Lucia, Queensland 4067, Australia}

\author{Leo Sementilli}
\affiliation{School of Mathematics and Physics, The University of Queensland, St Lucia, Queensland 4067, Australia}
\affiliation{ARC Centre of Excellence for Engineered Quantum Systems, St Lucia, Queensland 4067, Australia}

\author{Victor M. Valenzuela}
\affiliation{School of Mathematics and Physics, The University of Queensland, St Lucia, Queensland 4067, Australia}
\affiliation{ARC Centre of Excellence for Engineered Quantum Systems, St Lucia, Queensland 4067, Australia}
\affiliation{Facultad de Ciencias Físico-Matemáticas, Universidad Autónoma de Sinaloa, Culiacan, Sinaloa, 80013, Mexico}

\author{Erick Romero}
\affiliation{School of Mathematics and Physics, The University of Queensland, St Lucia, Queensland 4067, Australia}
\affiliation{ARC Centre of Excellence for Engineered Quantum Systems, St Lucia, Queensland 4067, Australia}

\author{Glen I. Harris}
\affiliation{School of Mathematics and Physics, The University of Queensland, St Lucia, Queensland 4067, Australia}
\affiliation{ARC Centre of Excellence for Engineered Quantum Systems, St Lucia, Queensland 4067, Australia}

\author{Margaret Wegener}
\affiliation{School of Mathematics and Physics, The University of Queensland, St Lucia, Queensland 4067, Australia}

\author{Warwick P. Bowen}
\email{w.bowen@uq.edu.au}
\affiliation{School of Mathematics and Physics, The University of Queensland, St Lucia, Queensland 4067, Australia}
\affiliation{ARC Centre of Excellence for Engineered Quantum Systems, St Lucia, Queensland 4067, Australia}
\affiliation{ARC Centre of Excellence in Quantum Biotechnology, St. Lucia, Queensland 4067, Australia}

\date{\today}

\begin{abstract}
Accelerometers offer motion sensing capabilities across a wide range of areas, enabling navigational awareness in consumer goods and defense applications, and playing a key role in monitoring and control systems. To date, on-chip accelerometers have largely utilized a single device layer or substrate as a test mass. This constrains the test mass to the dimensions and density of the device layer or substrate, ultimately limiting the sensitivity of the device. We demonstrate a new approach which utilizes a pick-and-place bonding technique to increase the test mass of an on-chip accelerometer. By bonding a high-density platinum sphere to a nanomechanical silicon nitride trampoline membrane, we achieve a quality factor of 1900 in air with 95\,mg test mass, corresponding to a thermomechanical noise limited acceleration sensitivity of 0.8\,n\grtHz. We optically probe the device's response to applied accelerations with increasing level of acoustic and vibration isolation, measuring a peak sensitivity of 5.5\,n\grtHz\ at 117\,Hz in air, limited by environmental vibrations. This represents the best peak sensitivity reported using a chip-integrated test mass.
\end{abstract}

\maketitle

\section{Introduction}
Chip-scale micro-electromechanical accelerometers have enabled the widespread integration of motion sensing into modern consumer electronics, offering compact size, low power consumption, and scalable fabrication \cite{langfelder2021silicon}. These sensors are now ubiquitous in applications such as smartphones and wearable devices. Although they continue to improve, their suitability for high-precision applications—such as autonomous navigation and geophysical sensing—is limited by factors including thermal noise, long-term bias instability, and scale-factor drift \cite{ru2022mems}. Optomechanical sensors offer the potential to overcome these limitations, benefiting from resonant enhancement of motion and precise optical readout \cite{li2021cavity}. These advantages have seen optomechanical sensors applied in various types of sensing, including force \cite{fogliano2021ultrasensitive, melcher2014self, harris2013minimum}, acoustic \cite{basiri2019precision, cao2025ultrasonicsensors, wang2024high, mcqueen2025fibre}, pressure \cite{chen2022nano}, and magnetic \cite{forstner2012cavity, forstner2014ultrasensitive, hu2024picotesla, xu2024subpicotesla}. Optomechanical accelerometers have emerged as a promising alternative to their micro-electromechanical counterparts, leveraging the coupling between mechanical motion and high-quality optical cavities to achieve exquisite displacement sensitivity \cite{krause_high-resolution_2012}. This has enabled broadband readout with performance at the fundamental noise limits \cite{zhou2021broadband}. Recent work has demonstrated significant progress toward navigation-grade and deployable devices \cite{ge2025navigation}, paving the way for the next generation of high-performance on-chip accelerometers.

Acceleration sensing relies on precise displacement measurements of a test mass, which behaves in a similar way to a mass on a spring. While different detection schemes can be utilized, such as the aforementioned electrical and optical techniques, the ultimate limit of sensitivity is determined by the Brownian motion of the test mass. This thermal acceleration, $a_\text{th}$, is given by 
\begin{equation}
\label{eq:ath}
a_\text{th}=\sqrt{\frac{4 k_B T \omega_m}{Q m}},
\end{equation}
where $k_B$ is Boltzmann's constant, $T$ is the temperature, $\omega_m$ is the angular frequency of the mechanical resonance with quality factor $Q$, and $m$ is the mass \cite{krause_high-resolution_2012}. Therefore, for sensing at ambient temperature, the sensitivity depends on only two degrees of freedom: the quality factor and the mass. Increasing either of these parameters improves the sensitivity with square-root proportionality. 

For on-chip accelerometers, it is common to use the device layer or substrate itself as the test mass \cite{chowdhury2023membrane, li2024broadband, huang2020chip, qu2022passive}. This imposes a practical limit on the test mass based on the density and size of the chip layers. We overcome this constraint through hybrid integration of high-density proof masses with nanomechanical trampolines. We show the ability to functionalize silicon nitride nanomechanical structures with millimeter-scale platinum spheres using a pick-and-place technique. With a density nearly an order of magnitude higher than that of silicon, the use of platinum allows our nanomechanical accelerometer to operate with a test mass twice that of previous devices \cite{qu2022passive}. The sensor achieves thermal noise limited sensitivity as low as 0.8\,n\grtHz\ ($g = 9.81\,\textrm{ms}^{-2}$, not to be confused with the gram, g). We tested the response of our devices to applied acceleration in air, taking measurements over many hours. We find that the high sensitivity requires the use of thorough acoustic and vibration shielding, and develop methods to achieve this. 
\begin{figure*}[htbp]
    \centering
    \includegraphics{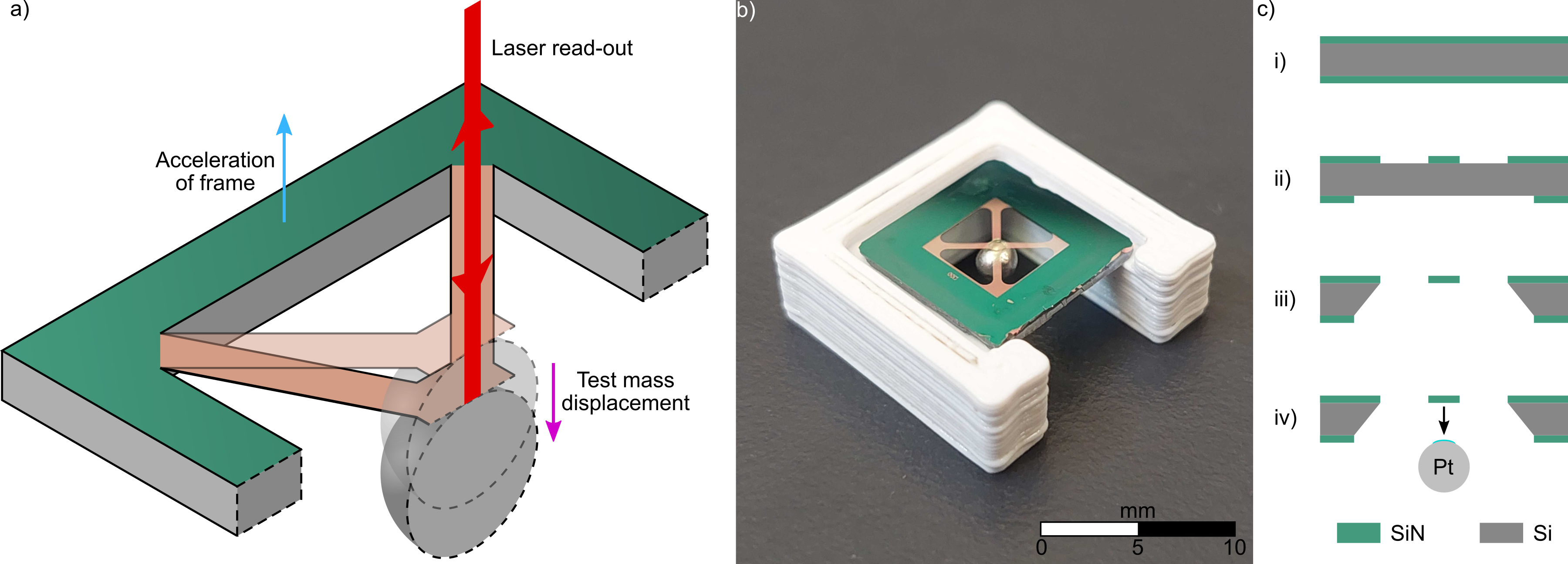}
    \caption{a) Accelerometer operating principle. b) Photograph of the assembled sensor. c) Fabrication workflow: i) low-stress SiN on Si chip; ii) photolithography followed by reactive-ion etching; iii) anisotropic KOH silicon etch; iv) hybrid test mass integration.}
    \label{fig:1}
\end{figure*}
This allows us to achieve a peak sensitivity of 5.5\,n\grtHz, a factor of seven above the thermal noise limit, limited by residual vibration noise. The devices demonstrated here have sensitivity a factor of six better than previously reported optomechanical accelerometers with chip-scale test masses \cite{zhou2021broadband}, opening new opportunities for acceleration sensing. They also have potential applications in fundamental physics, such as investigations into quantum gravity \cite{liu2021gravitational, tang2025cavityoptomechanicalprobegravity} and in laser interferometer gravitational wave observatories \cite{carter2024compact, hines2023compact}.

\section{Design and fabrication}
The operating principle of our device is shown in Figure~\ref{fig:1}a, where the inertia of the suspended test mass causes a displacement relative to the supporting frame when the system undergoes acceleration perpendicular to the frame. This displacement is measured using laser interferometry, in which the acceleration-induced motion of the test mass changes the optical path length. This results in a phase shift of the interferometric signal, which is directly proportional to the displacement and can be used to infer the applied acceleration. A fabricated device is shown in Figure~\ref{fig:1}b. The 500\,nm thick, low-stress silicon nitride (UniversityWafer) trampoline resonator consists of four 400\,\textmu m wide clamp-rounded tethers (250\,\textmu m radius of curvature) in a 5\,mm square frame that supports a 0.8\,mm $\times$ 0.8\,mm center trampoline pad. To fabricate these trampoline structures (Figure~\ref{fig:1}c), we first define the silicon nitride geometry using photolithography (Heidelberg Instruments MLA 150) and reactive-ion etching (Oxford PlasmaPro 100 Cobra ICP RIE). An aggressive $90^\circ$C anisotropic KOH silicon etch is then performed on the back side of the chip only, followed by a gentle $50^\circ$C KOH etch from both sides for the final release (both using 30\% KOH by weight). Before mass loading, the fabricated trampolines exhibit a mechanical quality factor $Q = 1.3\times10^{5}$ for the 13.8\,kHz fundamental out-of-plane mode when measured under vacuum ($<10^{-5}$\,mbar). This is lower than the quality factors achieved with highly stressed silicon nitride trampolines, which often exceed ten million \cite{engelsen2024ultrahigh, sementilli2022nanomechanical}, but is consistent with expectations for low-stress silicon nitride and low-frequency operation, where radiative losses can become significant \cite{Villanueva2014Evidence}.

A spherical 95\,mg platinum test mass with a radius of 1\,mm is attached to the trampoline resonator using a pick-and-place process. The resonator chip is lifted with a vacuum pick-up, aligned to the mass, and bonded using UV adhesive (Appendix~\ref{app:test mass bonding}). This process is highly repeatable. It allows us to overcome limitations imposed on the test mass size when it is patterned from the device substrate. The substrate material has relatively low density and, due to limited substrate thickness, large substrate-based test masses require substantially increased area \cite{qu2022passive}, constraining the total device footprint. For comparison, a mass equivalent to our platinum sphere, but composed of the 0.5\,mm thick silicon substrate, would be 9\,mm $\times$ 9\,mm---approximately 25 times larger in area. A further advantage that our technique provides is the control over the test mass properties such as shape, dimensions, and density. As a result, specific sensitivities and frequencies can be targeted without potential fabrication limitations, and particular materials can be chosen for specific applications. Platinum was chosen as the test mass material in this work due to its high density, availability and ease of handling. Devices were also successfully constructed with gold test masses, but resulted in lighter masses because of their slightly lower density. 

\section{Measurements and results}
\begin{figure*}[htbp]
    \centering
    \includegraphics{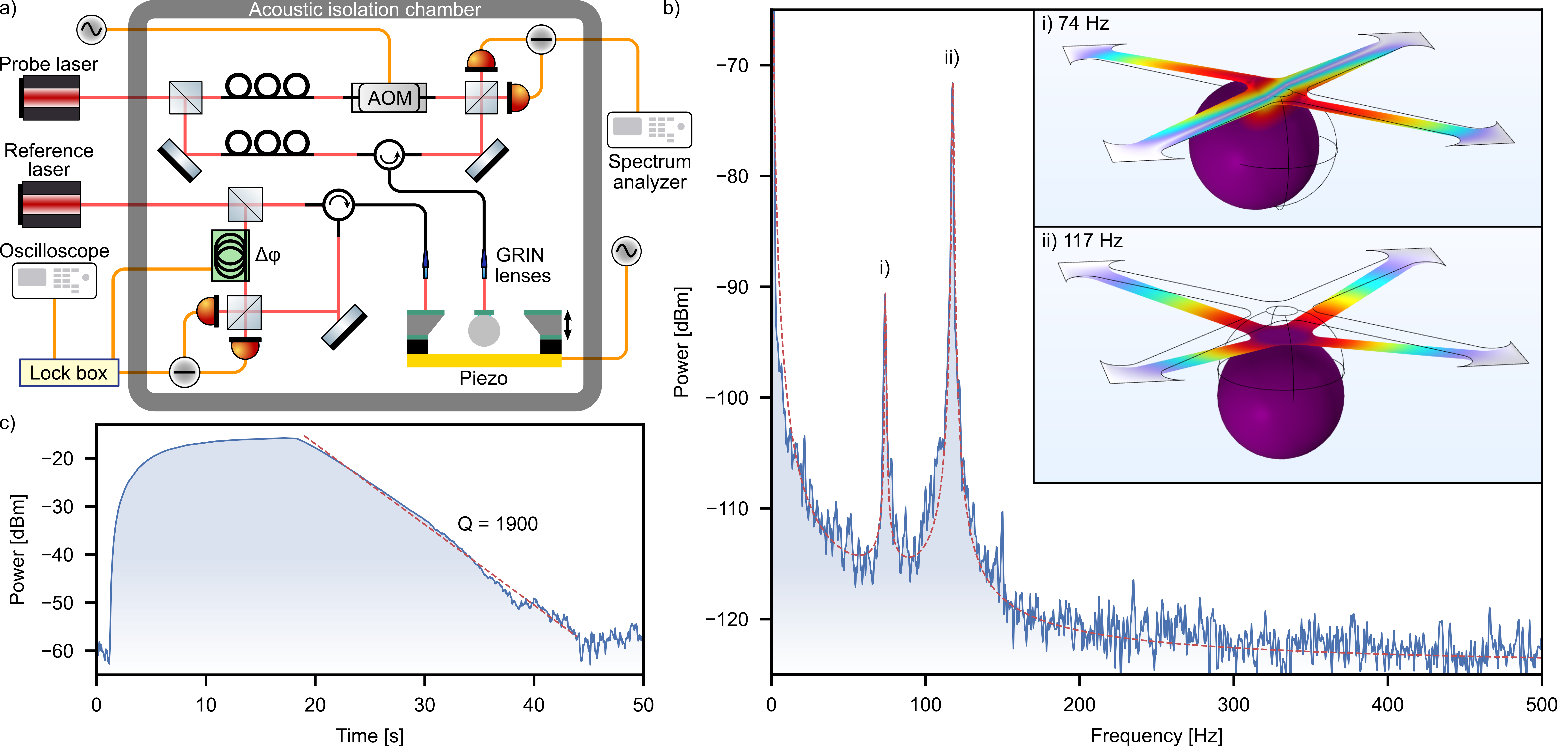}
    \caption{a) Schematic diagram of the measurement setup. AOM: acousto-optic modulator, $\Delta\varphi$: fiber stretcher. b) Measured power spectrum of the device. The red dashed line represents a model including a white noise drive of the mechanical resonator, thermomechanical noise, optical shot noise, and 1/f noise. Inset: finite element modeling of the i) rocking and ii) out-of-plane mechanical resonances with their corresponding frequencies. c) Ring down measurement in air with corresponding quality factor.}
    \label{fig:2}
\end{figure*}

The device was measured at atmospheric pressure using heterodyne interferometry (Figure~\ref{fig:2}a). 
For applied acceleration measurements, the device was placed on top of a piezoelectric actuator, with a separate homodyne measurement scheme used to calibrate the applied acceleration. The device and associated measurement interferometers were mounted on an active vibration isolation stage (Herzan TS-150), contained within an acoustic isolation chamber (Herzan Silencer), which was mounted on a Newport RS2000 optical table (610\,mm thick, tuned damping) with Newport S2000 laminar flow supports.

The frequency spectrum in Figure~\ref{fig:2}b shows the fundamental out-of-plane mode at 117\,Hz, alongside a lower frequency rocking mode at 74\,Hz. These modes are substantiated by finite element modeling using COMSOL. A ring-down measurement of the 117\,Hz out-of-plane mode gives a mechanical quality factor $Q = 1900$ (Figure~\ref{fig:2}c). The mechanical damping rate $\Gamma = \omega_m/Q$, relevant for Equation~\ref{eq:ath}, remains similar to that of the unloaded trampoline. We believe that the current design is limited by the bending and radiative losses at the clamping points and the effect of the adhesive. This could be improved by engineering the trampoline structure to reduce radiative losses and potentially leverage the effects of dissipation dilution through the use of a different resonator material \cite{engelsen2024ultrahigh, sementilli2022nanomechanical, bereyhi2022hierarchical} or by targeting the torsional mode \cite{pratt2023nanoscale, hyatt2025ultrahigh}. In addition, optimizing the mass bonding method and materials \cite{liu2021gravitational}, and measuring under vacuum to remove potential gas damping effects, could provide further improvement. Using Equation~(\ref{eq:ath}) and the current measured quality factor, we determine that the 117\,Hz mode has a thermal noise limited acceleration sensitivity of 0.8\,n\grtHz.

\begin{figure*}[htbp]
    \centering
    \includegraphics{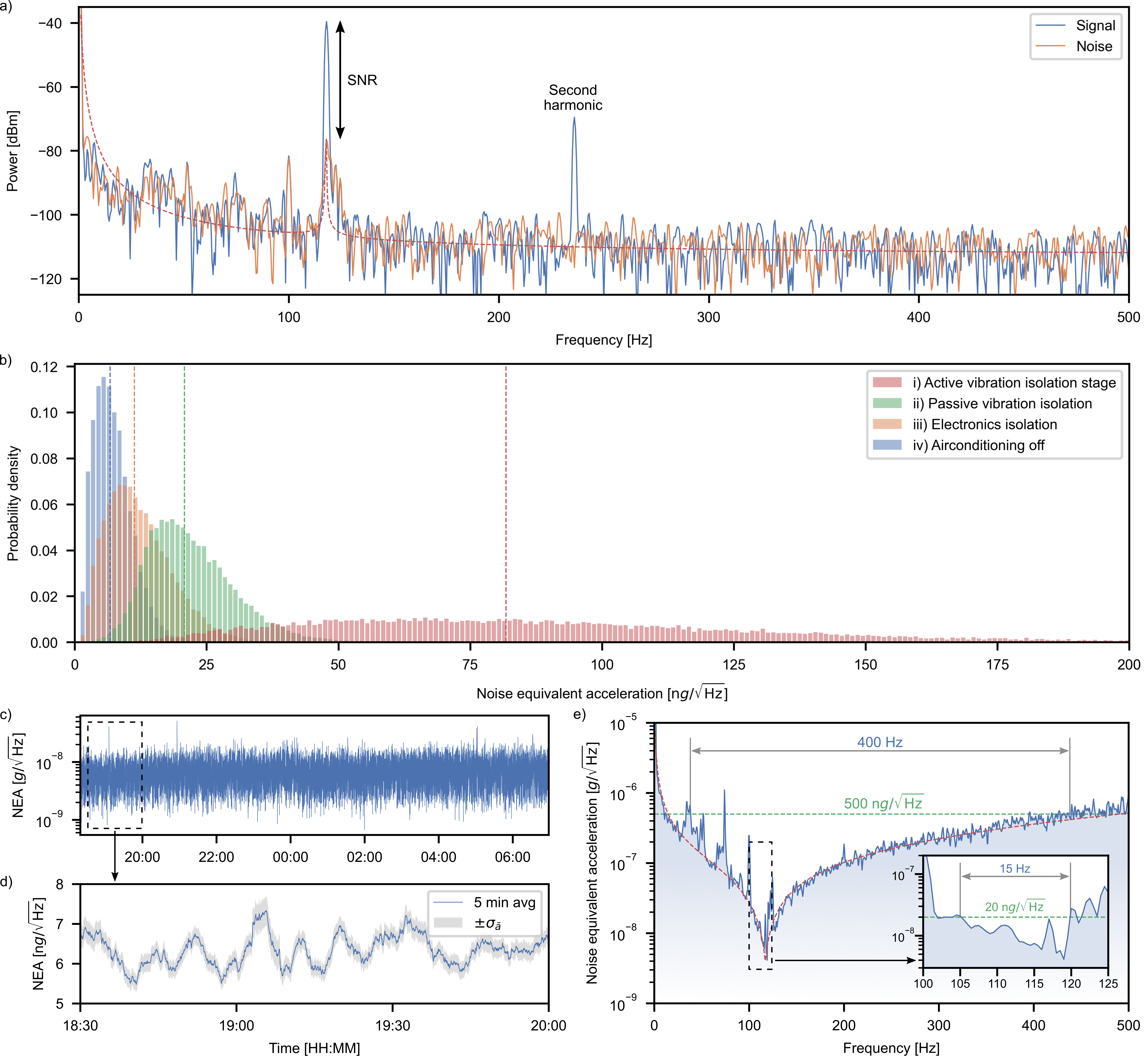}
    \caption{a) Power spectral density plot showing the noise floor and the response to an applied 390\,n$g$ acceleration at 117\,Hz, measured with a resolution bandwidth of 1\,Hz. The red dashed line represents a model including a white noise drive of the mechanical resonator, thermomechanical noise, optical shot noise, and 1/f noise. b) Histograms of the measured noise equivalent acceleration at the 117\,Hz resonance frequency for different levels of isolation. Dashed lines indicate the median values: i) Measurement in the acoustic isolation chamber. ii) Measurement with the acoustic isolation chamber on a floating optical bench. iii) Isolation of the control and measurement electronics from the optical bench and acoustic isolation chamber. iv) Laboratory air conditioning system turned off. c) Twelve-hour time trace of the lowest noise data (histogram iv). d) Ninety-minute trace of data averaged over five minutes, with the standard error of the mean represented by the shaded region. e) Noise equivalent acceleration as a function of frequency. Sub 500\,n\grtHz\ is achieved over a 400\,Hz bandwidth and sub 20\,n\grtHz\ is achieved over a 15\,Hz bandwidth. The red dashed line represents a model including a white noise drive of the mechanical resonator, thermomechanical noise, optical shot noise, and 1/f noise.}
    \label{fig:3}
\end{figure*}

The aim of accelerometry is to measure the inertial lag of the motion of the test mass relative to the device frame. The standard form of the equation of motion used for acceleration sensing is therefore given by
\begin{equation}
    m(\ddot{x_r} + \Gamma\dot{x_r} + \omega_m^{2}x_r) = -m\ddot{x_f},
\end{equation}
where $x_f$ is the displacement of the device frame, and $x_r$ is the relative displacement between the mass and the frame \cite{zhou2021broadband}. The noise equivalent acceleration can then be calculated in the same way as the sensitivity of other optomechanical sensors \cite{forstner2012cavity, mcqueen2025fibre},
\begin{equation}
    \text{NEA}(\omega) = \frac{a_\text{applied}(\omega)}{\sqrt{\text{SNR}(\omega) \times \text{RBW}}},
\end{equation}
where $a_\text{applied}$ is the acceleration applied at some frequency $\omega$, SNR is the signal-to-noise ratio measured in the sensor reference frame, and RBW is the resolution bandwidth of the measurement. In our case, we measure the position of the test mass in the lab frame rather than the accelerometer frame. To account for this, we make the substitution $x_t = x_r + x_f$, where $x_t$ is the total displacement we measure. The equation of motion then becomes
\begin{equation}
    m(\ddot{x_t} + \Gamma\dot{x_r} + \omega_m^2x_r) = 0.
\end{equation}
Applying the Fourier transform and taking the ratio of the relative to total displacement gives
\begin{equation}
    \frac{x_r(\omega)}{x_t(\omega)} = \frac{\omega^2}{\omega_m^2 + i\omega\Gamma}.
\end{equation}
This correction factor is applied to the measured signal in our calculation of the noise equivalent acceleration, which becomes
\begin{equation}
    \text{NEA}(\omega) = \frac{a_\text{applied}(\omega)}{\sqrt{\text{SNR}(\omega) \times \text{RBW}}}\sqrt{\frac{\omega_m^4 + \omega^2\Gamma^2}{\omega^4}}.
\end{equation}

To measure the noise equivalent acceleration of our device, we apply a sinusoidal drive to the piezoelectric actuator at the 117\,Hz mechanical resonance frequency. A strong response from the device is observed on a spectrum analyzer. It is shown and compared to the noise floor in Figure~\ref{fig:3}a. The displacement of the device frame is measured with the reference interferometer to determine the acceleration as a function of the drive voltage. This relationship is extrapolated to the low voltages used in measurements to estimate an applied root-mean-square acceleration of 390\,n$g$.

Initial measurements led to calculated noise equivalent accelerations over two orders of magnitude worse than expected from the thermal noise limit and with large variability. This was due to environmental noise driving the mechanical resonator. Measurement of the laboratory environment showed an acceleration noise at the 1\,\textmu $g$ level (Appendix~\ref{app:lab noise}). This corresponds to the vibration criterion level VC-J \cite{amick2005evolving}, an already extremely quiet environment, but still orders of magnitude above the expected limitation of the accelerometer. Efforts were then focused on isolating the sensor from ambient noise sources.

Measurements of the noise equivalent acceleration at the 117\,Hz resonant frequency were taken continuously overnight to identify periods of low environmental noise resulting in a median value of $81.8 \pm 0.4$\,n\grtHz\ with a standard deviation of $\sigma = 43$\,n\grtHz. This was repeated with the active vibration isolation stage turned off, which showed an improvement in the measured values to a median of $20.80 \pm 0.06$\,n\grtHz\ with $\sigma = 8.1$\,n\grtHz. We conclude that noise from the piezoelectric sensors and force-motors in the active vibration isolation stage was driving the mechanical motion of our device. The active vibration isolation stage was replaced by an optical breadboard mounted on a layer of passive vibration isolating polymer for subsequent measurements in which all the auxiliary electronics were further isolated from the system or moved outside the laboratory. This configuration reduced the median noise equivalent acceleration to $11.29 \pm 0.05$\,n\grtHz\ ($\sigma = 6.2$\,n\grtHz). Finally, the laboratory air conditioning system was turned off and the noise floor reached a median of $6.72 \pm 0.03$\,n\grtHz\ ($\sigma = 3.7$\,n\grtHz) over a full twelve-hour overnight measurement. These measurements are plotted as histograms in Figure~\ref{fig:3}b and show marked improvements in both noise equivalent acceleration and its variability.

Figure~\ref{fig:3}c shows a twelve-hour continuous measurement of the noise equivalent acceleration at 117\,Hz with the full vibration and acoustic isolation system in place. A ninety-minute period of this trace is shown in Figure~\ref{fig:3}d, with a five-minute averaging window applied. This reaches noise equivalent accelerations as low as 5.5\,n\grtHz, the lowest noise floor reported for an accelerometer with chip-scale test mass (Table \ref{tab:table1}), and approaching the sub-n$g$ performance of larger gram-scale devices \cite{carter2024compact}. Despite these improvements, the measured noise equivalent acceleration is still appreciably above the theoretical thermomechanical noise limit of 0.8\,n\grtHz. We suspect that this is due to residual environmental noise.

The frequency response of the accelerometer was characterised by sweeping the drive frequency and measuring the corresponding signal-to-noise ratio. Away from the mechanical resonance, it was necessary to increase the drive strength since there is little to no resonant enhancement of the sensor displacement. The frequency response of the driving piezoelectric actuator was measured using the same procedure to calibrate the applied acceleration over the frequency band. The resulting frequency response is shown in Figure~\ref{fig:3}e and is in good agreement with a model based on a white noise drive of the fundamental out-of-plane mechanical mode, thermomechanical noise, optical shot noise, and 1/f noise. Sensitivity below 20\,n\grtHz\ is achieved over a modest bandwidth of 15\,Hz, while sub 500\,n\grtHz\ sensitivity is achieved in the range from 40\,Hz to 440\,Hz (excluding the peak due to the rocking mode). The bandwidth could be further improved with the integration of optical cavity-based readout, allowing for broadband mechanical noise limited sensing \cite{li2021cavity}.

A comparison of key parameters for chip-based optomechanical accelerometers reported in the literature that use chip-scale test masses, is summarised in Table~\ref{tab:table1}. The flexibility of our approach allows for a large test mass, twice the mass of the next largest, enabling us to achieve the lowest noise equivalent acceleration reported to date. The modest quality factor and resonant frequency of our device can be increased through improvements to the design structure and choice of materials. Our use of heterodyne, rather than cavity-enhanced readout, degrades bandwidth. The increased test mass enables sub-500\,n\grtHz\ sensitivity over a 400\,Hz bandwidth, but still a factor of eighty less than the best that has previously been reported \cite{ge2025navigation}. The use of chip-integrated optical readout would improve the bandwidth of our sensor.

\begin{table*}[htbp]
\caption{Comparison of chip-based optomechanical accelerometers reported in the literature.}
\label{tab:table1}
\begin{ruledtabular}
\begin{tabular}{lccccc}
\textrm{Ref.}&
\textrm{$m$ [mg]}&
\textrm{NEA [\textmu\grtHz]}&
\textrm{$Q$}&
\textrm{$\omega_m/2\pi$ [kHz]}&
\textrm{BW $< 500$\,n\grtHz\ [kHz]}\\
\colrule
\cite{krause_high-resolution_2012}\footnote{Measured in vacuum} & $1\times10^{-5}$ & 10 & $1.4\times10^6$ & 27.5 & -\\
\cite{zhou2021broadband}\footnotemark[1]  & 20 & 0.032 & 565 & 8.74 & 27.5\\
\cite{ge2025navigation}  & 16.7 & 4.2 & 54.3 & 93.4 & 32\\
\cite{chowdhury2023membrane}\footnotemark[1] & $1.2\times10^{-5}$ & 0.6 & $1.1\times10^7$ & 40 & -\\
\cite{li2024broadband}\footnotemark[1]\footnote{$T$ = 50\,K} & $3.5\times10^{-4}$ & 0.72 & $4.1\times10^5$ & 412.59 & -\\
\cite{huang2020chip}\footnotemark[1]  & $5.6\times10^{-3}$ & 8.2 & 1383 & 71.3 & -\\
\cite{qu2022passive} & 44.1 & 0.067 & 225 & 0.113 & 0.1\\
This work & 95 & 0.0055 & 1900 & 0.117 & 0.4\\
\end{tabular}
\end{ruledtabular}
\end{table*}

\section{Conclusion}
We have developed an optomechanical accelerometer with a hybrid-integrated test mass, using a bonded platinum test mass and a silicon nitride membrane. This allows the test mass to be increased by a factor of two compared to previous on-chip optomechanical accelerometers. With this approach and employing extensive acoustic and vibration isolation, we achieved noise equivalent acceleration as low as 5.5\,n\grtHz\ in air.

Advances in optomechanical accelerometers have produced significant improvements in sensitivity, stability, and integration. These low-noise, low-drift devices are well suited for applications in GPS-denied navigation, enhanced gravitational mapping, and may allow new tests of fundamental physics. As performance and robustness continue to progress, optomechanical sensors are emerging as strong candidates for practical, field-deployable technologies. By providing new methods for hybrid test mass integration, our work provides a path to higher sensitivity and greater flexibility device designs, supporting this transition towards applications.

\begin{acknowledgments}
The authors acknowledge the facilities, and the scientific and technical assistance, of the Australian Microscopy $\&$ Microanalysis Research Facility at the Centre for Microscopy and Microanalysis, The University of Queensland. This work was performed in part at the Queensland node of the Australian National Fabrication Facility, a company established under the National Collaborative Research Infrastructure Strategy to provide nano and microfabrication facilities for Australia’s researchers. The Commonwealth of Australia (represented by the Defence Science and Technology Group) supports this research through a Defence Science Partnerships agreement. This work was funded under the NGTF and is being delivered through the ASCA. The authors thank Dr.\ Scott Foster (Defence Science and Technology Group) for his support, expertise, and discussions on the research in this publication. This work was also financially supported by the Australian Research Council Centre of Excellence for Engineered Quantum Systems (EQUS, Grant No. CE170100009) and the Australian Research Council Centre of Excellence in Quantum Biotechnology (QUBIC, Grant No. CE230100021). We thank the Queensland Defence Science Alliance (QDSA) for financial support of the project through the 2023 QDSA Collaborative Research Grant (CRG) funding round. This project was funded by the Queensland Government through the Department of Environment, Tourism, Science and Innovation’s (DETSI) Quantum 2032 Challenge Program. The program accelerates quantum sportstech, connects Queensland's research sector with industry, and showcases the state's quantum capabilities as part of Brisbane 2032's legacy. The authors thank Carlos Ramirez and Kevin Jack for performing the laboratory noise measurements. Victor M.\ Valenzuela also acknowledges support from SECIHTI (Scholarship No. 828253) and a sabbatical granted by the Universidad Autónoma de Sinaloa.
\end{acknowledgments}

\appendix
\section{Vacuum ring down}
\begin{figure}[htbp]
    \centering
    \includegraphics{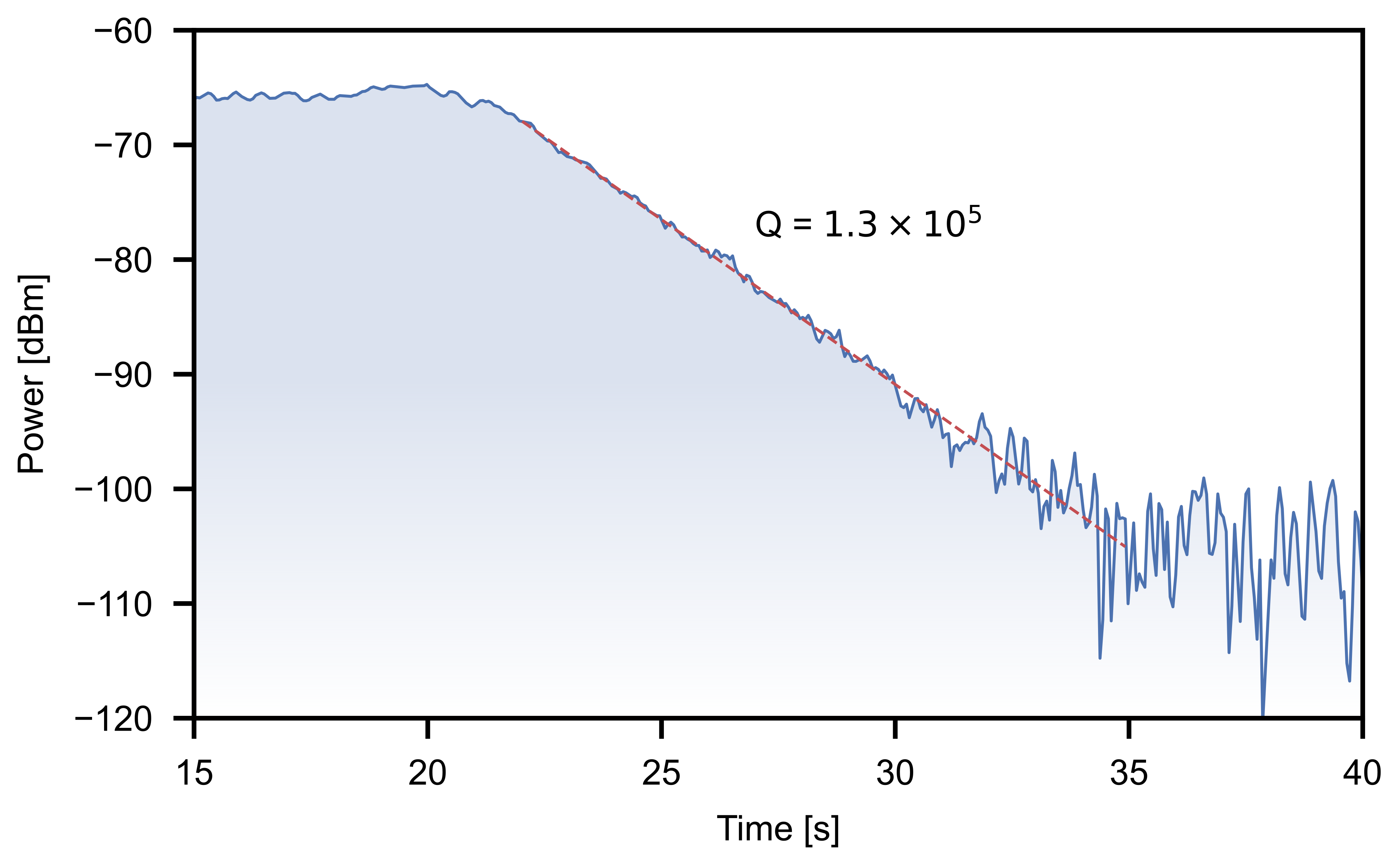}
    \caption{Mechanical ring down measurement of the silicon nitride trampoline in vacuum prior to test mass bonding.}
    \label{fig:VacRingdown}
\end{figure}

Before bonding the platinum test mass, the mechanical properties of the silicon nitride trampoline structures were measured in a vacuum chamber at a pressure less than $10^{-5}$\,mbar. A quality factor of $1.3\times10^{5}$ was obtained from the mechanical ring down measurement for the fundamental out-of-plane mode at 13.8\,kHz (Figure~\ref{fig:VacRingdown}).

\section{Test mass bonding} \label{app:test mass bonding}
\begin{figure}[htbp]
    \centering
    \includegraphics{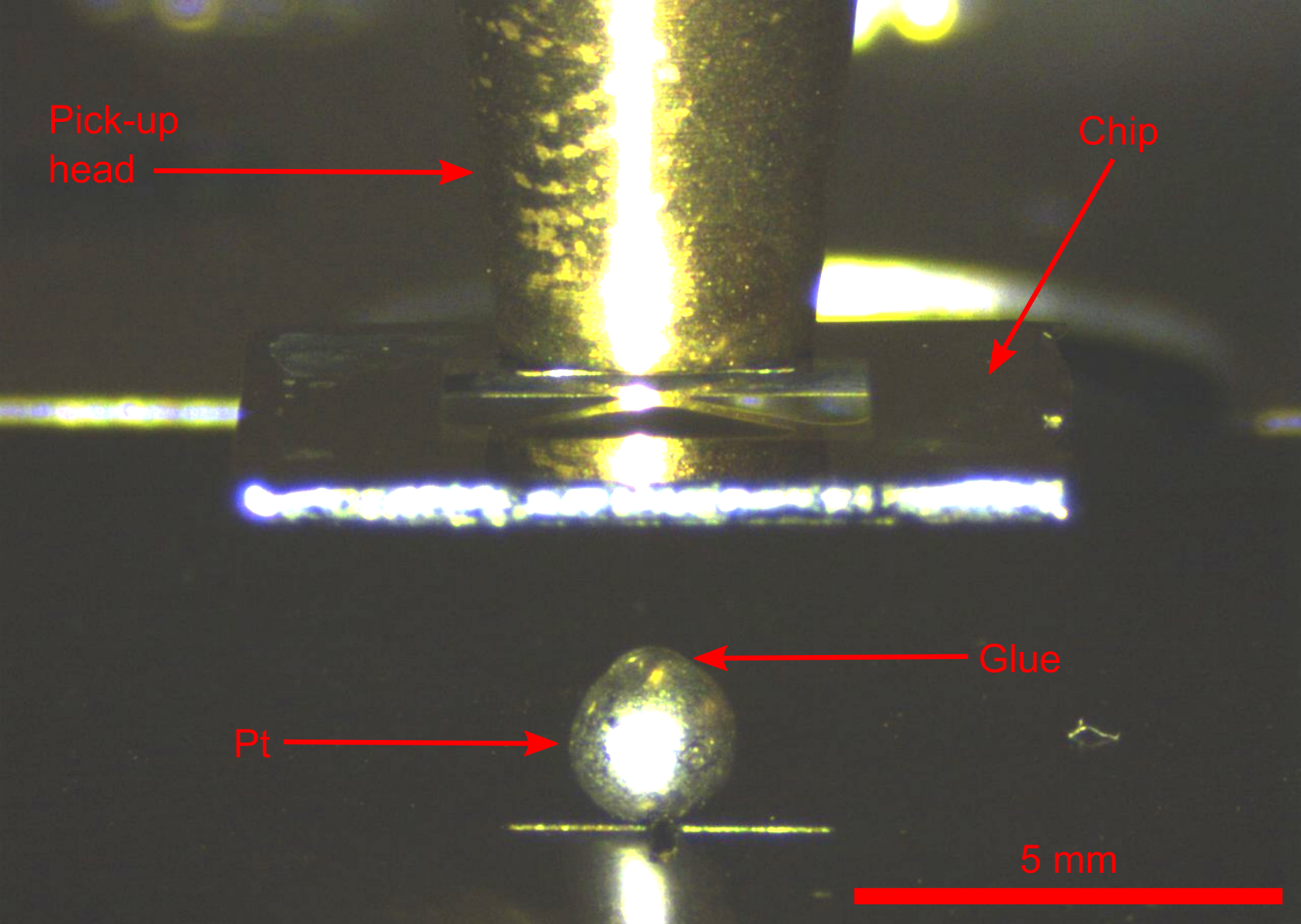}
    \caption{Photograph of the test mass bonding process.}
    \label{fig:5}
\end{figure}

Test mass bonding is carried out using a Finetech Fineplacer Lambda. We first lift the trampoline resonators using the vacuum pick-up head. The mass is then aligned to the trampoline resonator using the built-in beam splitter, camera, and translation stages. A small amount of UV glue (NOA 86H) is applied to the platinum mass, which is then brought into contact with the trampoline resonator using micromanipulation stages. Upon contact, the glue is cured using UV light. A photograph of the process is shown in Figure~\ref{fig:5}.

\section{Laboratory noise measurements} \label{app:lab noise}
\begin{figure}[htbp]
    \centering
    \includegraphics{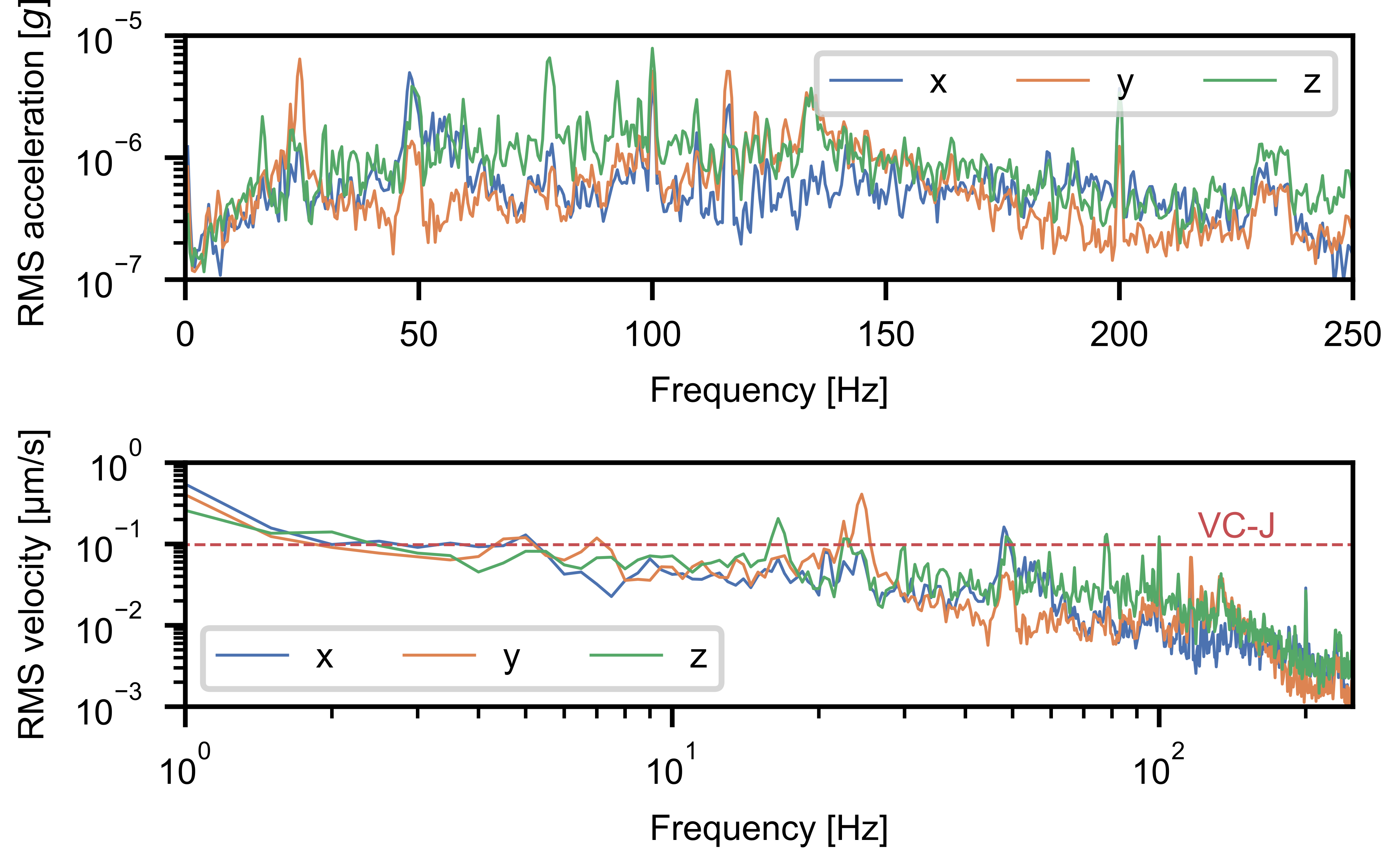}
    \caption{Measured vibrational noise in the laboratory. Vibration criterion level VC-J is indicated by the dashed line.}
    \label{fig:lab noise}
\end{figure}

Laboratory noise measurements were made using a Spicer Consulting SC11 Auto Survey System. The vibrational noise measured on the laboratory floor is shown in Figure~\ref{fig:lab noise}. 
In the 50\,Hz to 150\,Hz range, the acceleration in the vertical ($z$) direction is on the order of 1\,\textmu$g$, highlighting the need to thoroughly isolate the sensor from environmental noise in order to measure the thermomechanical motion of the sensor. 

\bibliography{OM_accelerometer}

\begin{thebibliography}{33}%
\makeatletter
\providecommand \@ifxundefined [1]{%
 \@ifx{#1\undefined}
}%
\providecommand \@ifnum [1]{%
 \ifnum #1\expandafter \@firstoftwo
 \else \expandafter \@secondoftwo
 \fi
}%
\providecommand \@ifx [1]{%
 \ifx #1\expandafter \@firstoftwo
 \else \expandafter \@secondoftwo
 \fi
}%
\providecommand \natexlab [1]{#1}%
\providecommand \enquote  [1]{``#1''}%
\providecommand \bibnamefont  [1]{#1}%
\providecommand \bibfnamefont [1]{#1}%
\providecommand \citenamefont [1]{#1}%
\providecommand \href@noop [0]{\@secondoftwo}%
\providecommand \href [0]{\begingroup \@sanitize@url \@href}%
\providecommand \@href[1]{\@@startlink{#1}\@@href}%
\providecommand \@@href[1]{\endgroup#1\@@endlink}%
\providecommand \@sanitize@url [0]{\catcode `\\12\catcode `\$12\catcode `\&12\catcode `\#12\catcode `\^12\catcode `\_12\catcode `\%12\relax}%
\providecommand \@@startlink[1]{}%
\providecommand \@@endlink[0]{}%
\providecommand \url  [0]{\begingroup\@sanitize@url \@url }%
\providecommand \@url [1]{\endgroup\@href {#1}{\urlprefix }}%
\providecommand \urlprefix  [0]{URL }%
\providecommand \Eprint [0]{\href }%
\providecommand \doibase [0]{https://doi.org/}%
\providecommand \selectlanguage [0]{\@gobble}%
\providecommand \bibinfo  [0]{\@secondoftwo}%
\providecommand \bibfield  [0]{\@secondoftwo}%
\providecommand \translation [1]{[#1]}%
\providecommand \BibitemOpen [0]{}%
\providecommand \bibitemStop [0]{}%
\providecommand \bibitemNoStop [0]{.\EOS\space}%
\providecommand \EOS [0]{\spacefactor3000\relax}%
\providecommand \BibitemShut  [1]{\csname bibitem#1\endcsname}%
\let\auto@bib@innerbib\@empty
%</preamble>
\bibitem [{\citenamefont {Langfelder}\ \emph {et~al.}(2021)\citenamefont {Langfelder}, \citenamefont {Bestetti},\ and\ \citenamefont {Gadola}}]{langfelder2021silicon}%
  \BibitemOpen
  \bibfield  {author} {\bibinfo {author} {\bibfnamefont {G.}~\bibnamefont {Langfelder}}, \bibinfo {author} {\bibfnamefont {M.}~\bibnamefont {Bestetti}},\ and\ \bibinfo {author} {\bibfnamefont {M.}~\bibnamefont {Gadola}},\ }\href@noop {} {\bibfield  {journal} {\bibinfo  {journal} {Journal of Micromechanics and Microengineering}\ }\textbf {\bibinfo {volume} {31}},\ \bibinfo {pages} {084002} (\bibinfo {year} {2021})}\BibitemShut {NoStop}%
\bibitem [{\citenamefont {Ru}\ \emph {et~al.}(2022)\citenamefont {Ru}, \citenamefont {Gu}, \citenamefont {Shang},\ and\ \citenamefont {Zhang}}]{ru2022mems}%
  \BibitemOpen
  \bibfield  {author} {\bibinfo {author} {\bibfnamefont {X.}~\bibnamefont {Ru}}, \bibinfo {author} {\bibfnamefont {N.}~\bibnamefont {Gu}}, \bibinfo {author} {\bibfnamefont {H.}~\bibnamefont {Shang}},\ and\ \bibinfo {author} {\bibfnamefont {H.}~\bibnamefont {Zhang}},\ }\href@noop {} {\bibfield  {journal} {\bibinfo  {journal} {Micromachines}\ }\textbf {\bibinfo {volume} {13}},\ \bibinfo {pages} {879} (\bibinfo {year} {2022})}\BibitemShut {NoStop}%
\bibitem [{\citenamefont {Li}\ \emph {et~al.}(2021)\citenamefont {Li}, \citenamefont {Ou}, \citenamefont {Lei},\ and\ \citenamefont {Liu}}]{li2021cavity}%
  \BibitemOpen
  \bibfield  {author} {\bibinfo {author} {\bibfnamefont {B.-B.}\ \bibnamefont {Li}}, \bibinfo {author} {\bibfnamefont {L.}~\bibnamefont {Ou}}, \bibinfo {author} {\bibfnamefont {Y.}~\bibnamefont {Lei}},\ and\ \bibinfo {author} {\bibfnamefont {Y.-C.}\ \bibnamefont {Liu}},\ }\href@noop {} {\bibfield  {journal} {\bibinfo  {journal} {Nanophotonics}\ }\textbf {\bibinfo {volume} {10}},\ \bibinfo {pages} {2799} (\bibinfo {year} {2021})}\BibitemShut {NoStop}%
\bibitem [{\citenamefont {Fogliano}\ \emph {et~al.}(2021)\citenamefont {Fogliano}, \citenamefont {Besga}, \citenamefont {Reigue}, \citenamefont {Mercier~de L{\'e}pinay}, \citenamefont {Heringlake}, \citenamefont {Gouriou}, \citenamefont {Eyraud}, \citenamefont {Wernsdorfer}, \citenamefont {Pigeau},\ and\ \citenamefont {Arcizet}}]{fogliano2021ultrasensitive}%
  \BibitemOpen
  \bibfield  {author} {\bibinfo {author} {\bibfnamefont {F.}~\bibnamefont {Fogliano}}, \bibinfo {author} {\bibfnamefont {B.}~\bibnamefont {Besga}}, \bibinfo {author} {\bibfnamefont {A.}~\bibnamefont {Reigue}}, \bibinfo {author} {\bibfnamefont {L.}~\bibnamefont {Mercier~de L{\'e}pinay}}, \bibinfo {author} {\bibfnamefont {P.}~\bibnamefont {Heringlake}}, \bibinfo {author} {\bibfnamefont {C.}~\bibnamefont {Gouriou}}, \bibinfo {author} {\bibfnamefont {E.}~\bibnamefont {Eyraud}}, \bibinfo {author} {\bibfnamefont {W.}~\bibnamefont {Wernsdorfer}}, \bibinfo {author} {\bibfnamefont {B.}~\bibnamefont {Pigeau}},\ and\ \bibinfo {author} {\bibfnamefont {O.}~\bibnamefont {Arcizet}},\ }\href@noop {} {\bibfield  {journal} {\bibinfo  {journal} {Nature Communications}\ }\textbf {\bibinfo {volume} {12}},\ \bibinfo {pages} {4124} (\bibinfo {year} {2021})}\BibitemShut {NoStop}%
\bibitem [{\citenamefont {Melcher}\ \emph {et~al.}(2014)\citenamefont {Melcher}, \citenamefont {Stirling}, \citenamefont {Cervantes}, \citenamefont {Pratt},\ and\ \citenamefont {Shaw}}]{melcher2014self}%
  \BibitemOpen
  \bibfield  {author} {\bibinfo {author} {\bibfnamefont {J.}~\bibnamefont {Melcher}}, \bibinfo {author} {\bibfnamefont {J.}~\bibnamefont {Stirling}}, \bibinfo {author} {\bibfnamefont {F.~G.}\ \bibnamefont {Cervantes}}, \bibinfo {author} {\bibfnamefont {J.~R.}\ \bibnamefont {Pratt}},\ and\ \bibinfo {author} {\bibfnamefont {G.~A.}\ \bibnamefont {Shaw}},\ }\href@noop {} {\bibfield  {journal} {\bibinfo  {journal} {Applied Physics Letters}\ }\textbf {\bibinfo {volume} {105}} (\bibinfo {year} {2014})}\BibitemShut {NoStop}%
\bibitem [{\citenamefont {Harris}\ \emph {et~al.}(2013)\citenamefont {Harris}, \citenamefont {McAuslan}, \citenamefont {Stace}, \citenamefont {Doherty},\ and\ \citenamefont {Bowen}}]{harris2013minimum}%
  \BibitemOpen
  \bibfield  {author} {\bibinfo {author} {\bibfnamefont {G.~I.}\ \bibnamefont {Harris}}, \bibinfo {author} {\bibfnamefont {D.~L.}\ \bibnamefont {McAuslan}}, \bibinfo {author} {\bibfnamefont {T.~M.}\ \bibnamefont {Stace}}, \bibinfo {author} {\bibfnamefont {A.~C.}\ \bibnamefont {Doherty}},\ and\ \bibinfo {author} {\bibfnamefont {W.~P.}\ \bibnamefont {Bowen}},\ }\href@noop {} {\bibfield  {journal} {\bibinfo  {journal} {Physical review letters}\ }\textbf {\bibinfo {volume} {111}},\ \bibinfo {pages} {103603} (\bibinfo {year} {2013})}\BibitemShut {NoStop}%
\bibitem [{\citenamefont {Basiri-Esfahani}\ \emph {et~al.}(2019)\citenamefont {Basiri-Esfahani}, \citenamefont {Armin}, \citenamefont {Forstner},\ and\ \citenamefont {Bowen}}]{basiri2019precision}%
  \BibitemOpen
  \bibfield  {author} {\bibinfo {author} {\bibfnamefont {S.}~\bibnamefont {Basiri-Esfahani}}, \bibinfo {author} {\bibfnamefont {A.}~\bibnamefont {Armin}}, \bibinfo {author} {\bibfnamefont {S.}~\bibnamefont {Forstner}},\ and\ \bibinfo {author} {\bibfnamefont {W.~P.}\ \bibnamefont {Bowen}},\ }\href@noop {} {\bibfield  {journal} {\bibinfo  {journal} {Nature Communications}\ }\textbf {\bibinfo {volume} {10}},\ \bibinfo {pages} {132} (\bibinfo {year} {2019})}\BibitemShut {NoStop}%
\bibitem [{\citenamefont {Cao}\ \emph {et~al.}(2025)\citenamefont {Cao}, \citenamefont {Yang}, \citenamefont {Wang}, \citenamefont {Hu}, \citenamefont {Wu}, \citenamefont {Wang}, \citenamefont {Liu}, \citenamefont {Zhou}, \citenamefont {Li}, \citenamefont {Lao}, \citenamefont {Yang},\ and\ \citenamefont {Li}}]{cao2025ultrasonicsensors}%
  \BibitemOpen
  \bibfield  {author} {\bibinfo {author} {\bibfnamefont {X.}~\bibnamefont {Cao}}, \bibinfo {author} {\bibfnamefont {H.}~\bibnamefont {Yang}}, \bibinfo {author} {\bibfnamefont {M.}~\bibnamefont {Wang}}, \bibinfo {author} {\bibfnamefont {Z.-G.}\ \bibnamefont {Hu}}, \bibinfo {author} {\bibfnamefont {Z.-L.}\ \bibnamefont {Wu}}, \bibinfo {author} {\bibfnamefont {Y.}~\bibnamefont {Wang}}, \bibinfo {author} {\bibfnamefont {J.-F.}\ \bibnamefont {Liu}}, \bibinfo {author} {\bibfnamefont {X.}~\bibnamefont {Zhou}}, \bibinfo {author} {\bibfnamefont {J.}~\bibnamefont {Li}}, \bibinfo {author} {\bibfnamefont {C.}~\bibnamefont {Lao}}, \bibinfo {author} {\bibfnamefont {Q.-F.}\ \bibnamefont {Yang}},\ and\ \bibinfo {author} {\bibfnamefont {B.-B.}\ \bibnamefont {Li}},\ }\href {https://arxiv.org/abs/2506.20219} {\bibinfo {title} {Integrated optomechanical ultrasonic sensors with nano-pascal-level sensitivity}} (\bibinfo {year} {2025}),\ \Eprint {https://arxiv.org/abs/2506.20219} {arXiv:2506.20219 [physics.optics]} \BibitemShut
  {NoStop}%
\bibitem [{\citenamefont {Wang}\ \emph {et~al.}(2024)\citenamefont {Wang}, \citenamefont {Liu}, \citenamefont {Pan}, \citenamefont {Fan}, \citenamefont {Liu}, \citenamefont {Xing}, \citenamefont {Zhou}, \citenamefont {Tang},\ and\ \citenamefont {Liu}}]{wang2024high}%
  \BibitemOpen
  \bibfield  {author} {\bibinfo {author} {\bibfnamefont {R.}~\bibnamefont {Wang}}, \bibinfo {author} {\bibfnamefont {W.}~\bibnamefont {Liu}}, \bibinfo {author} {\bibfnamefont {Z.}~\bibnamefont {Pan}}, \bibinfo {author} {\bibfnamefont {W.}~\bibnamefont {Fan}}, \bibinfo {author} {\bibfnamefont {L.}~\bibnamefont {Liu}}, \bibinfo {author} {\bibfnamefont {E.}~\bibnamefont {Xing}}, \bibinfo {author} {\bibfnamefont {Y.}~\bibnamefont {Zhou}}, \bibinfo {author} {\bibfnamefont {J.}~\bibnamefont {Tang}},\ and\ \bibinfo {author} {\bibfnamefont {J.}~\bibnamefont {Liu}},\ }\href@noop {} {\bibfield  {journal} {\bibinfo  {journal} {Optics Express}\ }\textbf {\bibinfo {volume} {32}},\ \bibinfo {pages} {4816} (\bibinfo {year} {2024})}\BibitemShut {NoStop}%
\bibitem [{\citenamefont {McQueen}\ \emph {et~al.}(2025)\citenamefont {McQueen}, \citenamefont {Bawden}, \citenamefont {Carey}, \citenamefont {Marinkovi{\'c}}, \citenamefont {Bowen},\ and\ \citenamefont {Harris}}]{mcqueen2025fibre}%
  \BibitemOpen
  \bibfield  {author} {\bibinfo {author} {\bibfnamefont {L.~R.}\ \bibnamefont {McQueen}}, \bibinfo {author} {\bibfnamefont {N.}~\bibnamefont {Bawden}}, \bibinfo {author} {\bibfnamefont {B.~J.}\ \bibnamefont {Carey}}, \bibinfo {author} {\bibfnamefont {I.}~\bibnamefont {Marinkovi{\'c}}}, \bibinfo {author} {\bibfnamefont {W.~P.}\ \bibnamefont {Bowen}},\ and\ \bibinfo {author} {\bibfnamefont {G.~I.}\ \bibnamefont {Harris}},\ }\href@noop {} {\bibfield  {journal} {\bibinfo  {journal} {Optics Express}\ }\textbf {\bibinfo {volume} {33}},\ \bibinfo {pages} {25910} (\bibinfo {year} {2025})}\BibitemShut {NoStop}%
\bibitem [{\citenamefont {Chen}\ \emph {et~al.}(2022)\citenamefont {Chen}, \citenamefont {Liu}, \citenamefont {Hong}, \citenamefont {Zou}, \citenamefont {Liu}, \citenamefont {Luo},\ and\ \citenamefont {Wang}}]{chen2022nano}%
  \BibitemOpen
  \bibfield  {author} {\bibinfo {author} {\bibfnamefont {Y.}~\bibnamefont {Chen}}, \bibinfo {author} {\bibfnamefont {S.}~\bibnamefont {Liu}}, \bibinfo {author} {\bibfnamefont {G.}~\bibnamefont {Hong}}, \bibinfo {author} {\bibfnamefont {M.}~\bibnamefont {Zou}}, \bibinfo {author} {\bibfnamefont {B.}~\bibnamefont {Liu}}, \bibinfo {author} {\bibfnamefont {J.}~\bibnamefont {Luo}},\ and\ \bibinfo {author} {\bibfnamefont {Y.}~\bibnamefont {Wang}},\ }\href@noop {} {\bibfield  {journal} {\bibinfo  {journal} {ACS Applied Materials \& Interfaces}\ }\textbf {\bibinfo {volume} {14}},\ \bibinfo {pages} {39211} (\bibinfo {year} {2022})}\BibitemShut {NoStop}%
\bibitem [{\citenamefont {Forstner}\ \emph {et~al.}(2012)\citenamefont {Forstner}, \citenamefont {Prams}, \citenamefont {Knittel}, \citenamefont {Van~Ooijen}, \citenamefont {Swaim}, \citenamefont {Harris}, \citenamefont {Szorkovszky}, \citenamefont {Bowen},\ and\ \citenamefont {Rubinsztein-Dunlop}}]{forstner2012cavity}%
  \BibitemOpen
  \bibfield  {author} {\bibinfo {author} {\bibfnamefont {S.}~\bibnamefont {Forstner}}, \bibinfo {author} {\bibfnamefont {S.}~\bibnamefont {Prams}}, \bibinfo {author} {\bibfnamefont {J.}~\bibnamefont {Knittel}}, \bibinfo {author} {\bibfnamefont {E.}~\bibnamefont {Van~Ooijen}}, \bibinfo {author} {\bibfnamefont {J.}~\bibnamefont {Swaim}}, \bibinfo {author} {\bibfnamefont {G.}~\bibnamefont {Harris}}, \bibinfo {author} {\bibfnamefont {A.}~\bibnamefont {Szorkovszky}}, \bibinfo {author} {\bibfnamefont {W.}~\bibnamefont {Bowen}},\ and\ \bibinfo {author} {\bibfnamefont {H.}~\bibnamefont {Rubinsztein-Dunlop}},\ }\href@noop {} {\bibfield  {journal} {\bibinfo  {journal} {Physical review letters}\ }\textbf {\bibinfo {volume} {108}},\ \bibinfo {pages} {120801} (\bibinfo {year} {2012})}\BibitemShut {NoStop}%
\bibitem [{\citenamefont {Forstner}\ \emph {et~al.}(2014)\citenamefont {Forstner}, \citenamefont {Sheridan}, \citenamefont {Knittel}, \citenamefont {Humphreys}, \citenamefont {Brawley}, \citenamefont {Rubinsztein-Dunlop},\ and\ \citenamefont {Bowen}}]{forstner2014ultrasensitive}%
  \BibitemOpen
  \bibfield  {author} {\bibinfo {author} {\bibfnamefont {S.}~\bibnamefont {Forstner}}, \bibinfo {author} {\bibfnamefont {E.}~\bibnamefont {Sheridan}}, \bibinfo {author} {\bibfnamefont {J.}~\bibnamefont {Knittel}}, \bibinfo {author} {\bibfnamefont {C.~L.}\ \bibnamefont {Humphreys}}, \bibinfo {author} {\bibfnamefont {G.~A.}\ \bibnamefont {Brawley}}, \bibinfo {author} {\bibfnamefont {H.}~\bibnamefont {Rubinsztein-Dunlop}},\ and\ \bibinfo {author} {\bibfnamefont {W.~P.}\ \bibnamefont {Bowen}},\ }\href@noop {} {\bibfield  {journal} {\bibinfo  {journal} {Advanced materials (Deerfield Beach, Fla.)}\ }\textbf {\bibinfo {volume} {26}},\ \bibinfo {pages} {6348} (\bibinfo {year} {2014})}\BibitemShut {NoStop}%
\bibitem [{\citenamefont {Hu}\ \emph {et~al.}(2024)\citenamefont {Hu}, \citenamefont {Gao}, \citenamefont {Liu}, \citenamefont {Yang}, \citenamefont {Wang}, \citenamefont {Lei}, \citenamefont {Zhou}, \citenamefont {Li}, \citenamefont {Cao}, \citenamefont {Liang} \emph {et~al.}}]{hu2024picotesla}%
  \BibitemOpen
  \bibfield  {author} {\bibinfo {author} {\bibfnamefont {Z.-G.}\ \bibnamefont {Hu}}, \bibinfo {author} {\bibfnamefont {Y.-M.}\ \bibnamefont {Gao}}, \bibinfo {author} {\bibfnamefont {J.-F.}\ \bibnamefont {Liu}}, \bibinfo {author} {\bibfnamefont {H.}~\bibnamefont {Yang}}, \bibinfo {author} {\bibfnamefont {M.}~\bibnamefont {Wang}}, \bibinfo {author} {\bibfnamefont {Y.}~\bibnamefont {Lei}}, \bibinfo {author} {\bibfnamefont {X.}~\bibnamefont {Zhou}}, \bibinfo {author} {\bibfnamefont {J.}~\bibnamefont {Li}}, \bibinfo {author} {\bibfnamefont {X.}~\bibnamefont {Cao}}, \bibinfo {author} {\bibfnamefont {J.}~\bibnamefont {Liang}}, \emph {et~al.},\ }\href@noop {} {\bibfield  {journal} {\bibinfo  {journal} {Light: Science \& Applications}\ }\textbf {\bibinfo {volume} {13}},\ \bibinfo {pages} {279} (\bibinfo {year} {2024})}\BibitemShut {NoStop}%
\bibitem [{\citenamefont {Xu}\ \emph {et~al.}(2024)\citenamefont {Xu}, \citenamefont {Li}, \citenamefont {Li}, \citenamefont {Liu},\ and\ \citenamefont {Liu}}]{xu2024subpicotesla}%
  \BibitemOpen
  \bibfield  {author} {\bibinfo {author} {\bibfnamefont {A.-N.}\ \bibnamefont {Xu}}, \bibinfo {author} {\bibfnamefont {Y.}~\bibnamefont {Li}}, \bibinfo {author} {\bibfnamefont {X.}~\bibnamefont {Li}}, \bibinfo {author} {\bibfnamefont {B.}~\bibnamefont {Liu}},\ and\ \bibinfo {author} {\bibfnamefont {Y.-C.}\ \bibnamefont {Liu}},\ }\href@noop {} {\bibfield  {journal} {\bibinfo  {journal} {Physical Review Letters}\ }\textbf {\bibinfo {volume} {133}},\ \bibinfo {pages} {153601} (\bibinfo {year} {2024})}\BibitemShut {NoStop}%
\bibitem [{\citenamefont {Krause}\ \emph {et~al.}(2012)\citenamefont {Krause}, \citenamefont {Winger}, \citenamefont {Blasius}, \citenamefont {Lin},\ and\ \citenamefont {Painter}}]{krause_high-resolution_2012}%
  \BibitemOpen
  \bibfield  {author} {\bibinfo {author} {\bibfnamefont {A.~G.}\ \bibnamefont {Krause}}, \bibinfo {author} {\bibfnamefont {M.}~\bibnamefont {Winger}}, \bibinfo {author} {\bibfnamefont {T.~D.}\ \bibnamefont {Blasius}}, \bibinfo {author} {\bibfnamefont {Q.}~\bibnamefont {Lin}},\ and\ \bibinfo {author} {\bibfnamefont {O.}~\bibnamefont {Painter}},\ }\href {https://doi.org/10.1038/nphoton.2012.245} {\bibfield  {journal} {\bibinfo  {journal} {Nature Photonics}\ }\textbf {\bibinfo {volume} {6}},\ \bibinfo {pages} {768} (\bibinfo {year} {2012})}\BibitemShut {NoStop}%
\bibitem [{\citenamefont {Zhou}\ \emph {et~al.}(2021)\citenamefont {Zhou}, \citenamefont {Bao}, \citenamefont {Madugani}, \citenamefont {Long}, \citenamefont {Gorman},\ and\ \citenamefont {LeBrun}}]{zhou2021broadband}%
  \BibitemOpen
  \bibfield  {author} {\bibinfo {author} {\bibfnamefont {F.}~\bibnamefont {Zhou}}, \bibinfo {author} {\bibfnamefont {Y.}~\bibnamefont {Bao}}, \bibinfo {author} {\bibfnamefont {R.}~\bibnamefont {Madugani}}, \bibinfo {author} {\bibfnamefont {D.~A.}\ \bibnamefont {Long}}, \bibinfo {author} {\bibfnamefont {J.~J.}\ \bibnamefont {Gorman}},\ and\ \bibinfo {author} {\bibfnamefont {T.~W.}\ \bibnamefont {LeBrun}},\ }\href@noop {} {\bibfield  {journal} {\bibinfo  {journal} {Optica}\ }\textbf {\bibinfo {volume} {8}},\ \bibinfo {pages} {350} (\bibinfo {year} {2021})}\BibitemShut {NoStop}%
\bibitem [{\citenamefont {Ge}\ \emph {et~al.}(2025)\citenamefont {Ge}, \citenamefont {Dominguez}, \citenamefont {Rubenok}, \citenamefont {Miller},\ and\ \citenamefont {Eichenfield}}]{ge2025navigation}%
  \BibitemOpen
  \bibfield  {author} {\bibinfo {author} {\bibfnamefont {C.}~\bibnamefont {Ge}}, \bibinfo {author} {\bibfnamefont {D.}~\bibnamefont {Dominguez}}, \bibinfo {author} {\bibfnamefont {A.}~\bibnamefont {Rubenok}}, \bibinfo {author} {\bibfnamefont {M.}~\bibnamefont {Miller}},\ and\ \bibinfo {author} {\bibfnamefont {M.}~\bibnamefont {Eichenfield}},\ }\href@noop {} {\bibinfo {title} {Towards navigation-grade and deployable optomechanical accelerometery}} (\bibinfo {year} {2025}),\ \Eprint {https://arxiv.org/abs/2505.11751} {arXiv:2505.11751} \BibitemShut {NoStop}%
\bibitem [{\citenamefont {Chowdhury}\ \emph {et~al.}(2023)\citenamefont {Chowdhury}, \citenamefont {Agrawal},\ and\ \citenamefont {Wilson}}]{chowdhury2023membrane}%
  \BibitemOpen
  \bibfield  {author} {\bibinfo {author} {\bibfnamefont {M.~D.}\ \bibnamefont {Chowdhury}}, \bibinfo {author} {\bibfnamefont {A.~R.}\ \bibnamefont {Agrawal}},\ and\ \bibinfo {author} {\bibfnamefont {D.~J.}\ \bibnamefont {Wilson}},\ }\href@noop {} {\bibfield  {journal} {\bibinfo  {journal} {Physical Review Applied}\ }\textbf {\bibinfo {volume} {19}},\ \bibinfo {pages} {024011} (\bibinfo {year} {2023})}\BibitemShut {NoStop}%
\bibitem [{\citenamefont {Li}\ \emph {et~al.}(2024)\citenamefont {Li}, \citenamefont {Liu}, \citenamefont {Liu}, \citenamefont {Gu}, \citenamefont {Liu}, \citenamefont {Zhou}, \citenamefont {Xing}, \citenamefont {Shi}, \citenamefont {Tang},\ and\ \citenamefont {Liu}}]{li2024broadband}%
  \BibitemOpen
  \bibfield  {author} {\bibinfo {author} {\bibfnamefont {W.}~\bibnamefont {Li}}, \bibinfo {author} {\bibfnamefont {W.}~\bibnamefont {Liu}}, \bibinfo {author} {\bibfnamefont {C.}~\bibnamefont {Liu}}, \bibinfo {author} {\bibfnamefont {Y.}~\bibnamefont {Gu}}, \bibinfo {author} {\bibfnamefont {L.}~\bibnamefont {Liu}}, \bibinfo {author} {\bibfnamefont {Y.}~\bibnamefont {Zhou}}, \bibinfo {author} {\bibfnamefont {E.}~\bibnamefont {Xing}}, \bibinfo {author} {\bibfnamefont {Y.}~\bibnamefont {Shi}}, \bibinfo {author} {\bibfnamefont {J.}~\bibnamefont {Tang}},\ and\ \bibinfo {author} {\bibfnamefont {J.}~\bibnamefont {Liu}},\ }\href@noop {} {\bibfield  {journal} {\bibinfo  {journal} {IEEE Sensors Journal}\ } (\bibinfo {year} {2024})}\BibitemShut {NoStop}%
\bibitem [{\citenamefont {Huang}\ \emph {et~al.}(2020)\citenamefont {Huang}, \citenamefont {Flor~Flores}, \citenamefont {Li}, \citenamefont {Wang}, \citenamefont {Wang}, \citenamefont {Goldberg}, \citenamefont {Zheng}, \citenamefont {Yu}, \citenamefont {Lu}, \citenamefont {Kutzer} \emph {et~al.}}]{huang2020chip}%
  \BibitemOpen
  \bibfield  {author} {\bibinfo {author} {\bibfnamefont {Y.}~\bibnamefont {Huang}}, \bibinfo {author} {\bibfnamefont {J.~G.}\ \bibnamefont {Flor~Flores}}, \bibinfo {author} {\bibfnamefont {Y.}~\bibnamefont {Li}}, \bibinfo {author} {\bibfnamefont {W.}~\bibnamefont {Wang}}, \bibinfo {author} {\bibfnamefont {D.}~\bibnamefont {Wang}}, \bibinfo {author} {\bibfnamefont {N.}~\bibnamefont {Goldberg}}, \bibinfo {author} {\bibfnamefont {J.}~\bibnamefont {Zheng}}, \bibinfo {author} {\bibfnamefont {M.}~\bibnamefont {Yu}}, \bibinfo {author} {\bibfnamefont {M.}~\bibnamefont {Lu}}, \bibinfo {author} {\bibfnamefont {M.}~\bibnamefont {Kutzer}}, \emph {et~al.},\ }\href@noop {} {\bibfield  {journal} {\bibinfo  {journal} {Laser \& Photonics Reviews}\ }\textbf {\bibinfo {volume} {14}},\ \bibinfo {pages} {1800329} (\bibinfo {year} {2020})}\BibitemShut {NoStop}%
\bibitem [{\citenamefont {Qu}\ \emph {et~al.}(2022)\citenamefont {Qu}, \citenamefont {Lu}, \citenamefont {Zhang}, \citenamefont {Liu},\ and\ \citenamefont {Zhang}}]{qu2022passive}%
  \BibitemOpen
  \bibfield  {author} {\bibinfo {author} {\bibfnamefont {Z.}~\bibnamefont {Qu}}, \bibinfo {author} {\bibfnamefont {P.}~\bibnamefont {Lu}}, \bibinfo {author} {\bibfnamefont {W.}~\bibnamefont {Zhang}}, \bibinfo {author} {\bibfnamefont {D.}~\bibnamefont {Liu}},\ and\ \bibinfo {author} {\bibfnamefont {J.}~\bibnamefont {Zhang}},\ }\href@noop {} {\bibfield  {journal} {\bibinfo  {journal} {Optics Express}\ }\textbf {\bibinfo {volume} {30}},\ \bibinfo {pages} {47587} (\bibinfo {year} {2022})}\BibitemShut {NoStop}%
\bibitem [{\citenamefont {Liu}\ \emph {et~al.}(2021)\citenamefont {Liu}, \citenamefont {Mummery}, \citenamefont {Zhou},\ and\ \citenamefont {Sillanp{\"a}{\"a}}}]{liu2021gravitational}%
  \BibitemOpen
  \bibfield  {author} {\bibinfo {author} {\bibfnamefont {Y.}~\bibnamefont {Liu}}, \bibinfo {author} {\bibfnamefont {J.}~\bibnamefont {Mummery}}, \bibinfo {author} {\bibfnamefont {J.}~\bibnamefont {Zhou}},\ and\ \bibinfo {author} {\bibfnamefont {M.~A.}\ \bibnamefont {Sillanp{\"a}{\"a}}},\ }\href@noop {} {\bibfield  {journal} {\bibinfo  {journal} {Physical Review Applied}\ }\textbf {\bibinfo {volume} {15}},\ \bibinfo {pages} {034004} (\bibinfo {year} {2021})}\BibitemShut {NoStop}%
\bibitem [{\citenamefont {Tang}\ \emph {et~al.}(2025)\citenamefont {Tang}, \citenamefont {Li}, \citenamefont {Sun}, \citenamefont {Cai}, \citenamefont {Li},\ and\ \citenamefont {Liu}}]{tang2025cavityoptomechanicalprobegravity}%
  \BibitemOpen
  \bibfield  {author} {\bibinfo {author} {\bibfnamefont {Z.}~\bibnamefont {Tang}}, \bibinfo {author} {\bibfnamefont {W.}~\bibnamefont {Li}}, \bibinfo {author} {\bibfnamefont {H.}~\bibnamefont {Sun}}, \bibinfo {author} {\bibfnamefont {X.}~\bibnamefont {Cai}}, \bibinfo {author} {\bibfnamefont {T.}~\bibnamefont {Li}},\ and\ \bibinfo {author} {\bibfnamefont {Y.}~\bibnamefont {Liu}},\ }\href {https://arxiv.org/abs/2506.13398} {\bibinfo {title} {Cavity optomechanical probe of gravity between massive mechanical oscillators}} (\bibinfo {year} {2025}),\ \Eprint {https://arxiv.org/abs/2506.13398} {arXiv:2506.13398 [quant-ph]} \BibitemShut {NoStop}%
\bibitem [{\citenamefont {Carter}\ \emph {et~al.}(2024)\citenamefont {Carter}, \citenamefont {Birckigt}, \citenamefont {Gerberding},\ and\ \citenamefont {Koehlenbeck}}]{carter2024compact}%
  \BibitemOpen
  \bibfield  {author} {\bibinfo {author} {\bibfnamefont {J.~J.}\ \bibnamefont {Carter}}, \bibinfo {author} {\bibfnamefont {P.}~\bibnamefont {Birckigt}}, \bibinfo {author} {\bibfnamefont {O.}~\bibnamefont {Gerberding}},\ and\ \bibinfo {author} {\bibfnamefont {S.~M.}\ \bibnamefont {Koehlenbeck}},\ }\href@noop {} {\bibfield  {journal} {\bibinfo  {journal} {Scientific Reports}\ }\textbf {\bibinfo {volume} {14}},\ \bibinfo {pages} {17775} (\bibinfo {year} {2024})}\BibitemShut {NoStop}%
\bibitem [{\citenamefont {Hines}\ \emph {et~al.}(2023)\citenamefont {Hines}, \citenamefont {Nelson}, \citenamefont {Zhang}, \citenamefont {Valdes}, \citenamefont {Sanjuan},\ and\ \citenamefont {Guzman}}]{hines2023compact}%
  \BibitemOpen
  \bibfield  {author} {\bibinfo {author} {\bibfnamefont {A.}~\bibnamefont {Hines}}, \bibinfo {author} {\bibfnamefont {A.}~\bibnamefont {Nelson}}, \bibinfo {author} {\bibfnamefont {Y.}~\bibnamefont {Zhang}}, \bibinfo {author} {\bibfnamefont {G.}~\bibnamefont {Valdes}}, \bibinfo {author} {\bibfnamefont {J.}~\bibnamefont {Sanjuan}},\ and\ \bibinfo {author} {\bibfnamefont {F.}~\bibnamefont {Guzman}},\ }\href@noop {} {\bibfield  {journal} {\bibinfo  {journal} {Applied Physics Letters}\ }\textbf {\bibinfo {volume} {122}} (\bibinfo {year} {2023})}\BibitemShut {NoStop}%
\bibitem [{\citenamefont {Engelsen}\ \emph {et~al.}(2024)\citenamefont {Engelsen}, \citenamefont {Beccari},\ and\ \citenamefont {Kippenberg}}]{engelsen2024ultrahigh}%
  \BibitemOpen
  \bibfield  {author} {\bibinfo {author} {\bibfnamefont {N.~J.}\ \bibnamefont {Engelsen}}, \bibinfo {author} {\bibfnamefont {A.}~\bibnamefont {Beccari}},\ and\ \bibinfo {author} {\bibfnamefont {T.~J.}\ \bibnamefont {Kippenberg}},\ }\href@noop {} {\bibfield  {journal} {\bibinfo  {journal} {Nature Nanotechnology}\ }\textbf {\bibinfo {volume} {19}},\ \bibinfo {pages} {725} (\bibinfo {year} {2024})}\BibitemShut {NoStop}%
\bibitem [{\citenamefont {Sementilli}\ \emph {et~al.}(2022)\citenamefont {Sementilli}, \citenamefont {Romero},\ and\ \citenamefont {Bowen}}]{sementilli2022nanomechanical}%
  \BibitemOpen
  \bibfield  {author} {\bibinfo {author} {\bibfnamefont {L.}~\bibnamefont {Sementilli}}, \bibinfo {author} {\bibfnamefont {E.}~\bibnamefont {Romero}},\ and\ \bibinfo {author} {\bibfnamefont {W.~P.}\ \bibnamefont {Bowen}},\ }\href@noop {} {\bibfield  {journal} {\bibinfo  {journal} {Advanced Functional Materials}\ }\textbf {\bibinfo {volume} {32}},\ \bibinfo {pages} {2105247} (\bibinfo {year} {2022})}\BibitemShut {NoStop}%
\bibitem [{\citenamefont {Villanueva}\ and\ \citenamefont {Schmid}(2014)}]{Villanueva2014Evidence}%
  \BibitemOpen
  \bibfield  {author} {\bibinfo {author} {\bibfnamefont {L.~G.}\ \bibnamefont {Villanueva}}\ and\ \bibinfo {author} {\bibfnamefont {S.}~\bibnamefont {Schmid}},\ }\href {https://doi.org/10.1103/PhysRevLett.113.227201} {\bibfield  {journal} {\bibinfo  {journal} {Phys. Rev. Lett.}\ }\textbf {\bibinfo {volume} {113}},\ \bibinfo {pages} {227201} (\bibinfo {year} {2014})}\BibitemShut {NoStop}%
\bibitem [{\citenamefont {Bereyhi}\ \emph {et~al.}(2022)\citenamefont {Bereyhi}, \citenamefont {Beccari}, \citenamefont {Groth}, \citenamefont {Fedorov}, \citenamefont {Arabmoheghi}, \citenamefont {Kippenberg},\ and\ \citenamefont {Engelsen}}]{bereyhi2022hierarchical}%
  \BibitemOpen
  \bibfield  {author} {\bibinfo {author} {\bibfnamefont {M.~J.}\ \bibnamefont {Bereyhi}}, \bibinfo {author} {\bibfnamefont {A.}~\bibnamefont {Beccari}}, \bibinfo {author} {\bibfnamefont {R.}~\bibnamefont {Groth}}, \bibinfo {author} {\bibfnamefont {S.~A.}\ \bibnamefont {Fedorov}}, \bibinfo {author} {\bibfnamefont {A.}~\bibnamefont {Arabmoheghi}}, \bibinfo {author} {\bibfnamefont {T.~J.}\ \bibnamefont {Kippenberg}},\ and\ \bibinfo {author} {\bibfnamefont {N.~J.}\ \bibnamefont {Engelsen}},\ }\href@noop {} {\bibfield  {journal} {\bibinfo  {journal} {Nature Communications}\ }\textbf {\bibinfo {volume} {13}},\ \bibinfo {pages} {3097} (\bibinfo {year} {2022})}\BibitemShut {NoStop}%
\bibitem [{\citenamefont {Pratt}\ \emph {et~al.}(2023)\citenamefont {Pratt}, \citenamefont {Agrawal}, \citenamefont {Condos}, \citenamefont {Pluchar}, \citenamefont {Schlamminger},\ and\ \citenamefont {Wilson}}]{pratt2023nanoscale}%
  \BibitemOpen
  \bibfield  {author} {\bibinfo {author} {\bibfnamefont {J.~R.}\ \bibnamefont {Pratt}}, \bibinfo {author} {\bibfnamefont {A.~R.}\ \bibnamefont {Agrawal}}, \bibinfo {author} {\bibfnamefont {C.~A.}\ \bibnamefont {Condos}}, \bibinfo {author} {\bibfnamefont {C.~M.}\ \bibnamefont {Pluchar}}, \bibinfo {author} {\bibfnamefont {S.}~\bibnamefont {Schlamminger}},\ and\ \bibinfo {author} {\bibfnamefont {D.~J.}\ \bibnamefont {Wilson}},\ }\href@noop {} {\bibfield  {journal} {\bibinfo  {journal} {Physical Review X}\ }\textbf {\bibinfo {volume} {13}},\ \bibinfo {pages} {011018} (\bibinfo {year} {2023})}\BibitemShut {NoStop}%
\bibitem [{\citenamefont {Hyatt}\ \emph {et~al.}(2025)\citenamefont {Hyatt}, \citenamefont {Agrawal}, \citenamefont {Pluchar}, \citenamefont {Condos},\ and\ \citenamefont {Wilson}}]{hyatt2025ultrahigh}%
  \BibitemOpen
  \bibfield  {author} {\bibinfo {author} {\bibfnamefont {A.~D.}\ \bibnamefont {Hyatt}}, \bibinfo {author} {\bibfnamefont {A.~R.}\ \bibnamefont {Agrawal}}, \bibinfo {author} {\bibfnamefont {C.~M.}\ \bibnamefont {Pluchar}}, \bibinfo {author} {\bibfnamefont {C.~A.}\ \bibnamefont {Condos}},\ and\ \bibinfo {author} {\bibfnamefont {D.~J.}\ \bibnamefont {Wilson}},\ }\href@noop {} {\bibfield  {journal} {\bibinfo  {journal} {arXiv preprint arXiv:2506.02325}\ } (\bibinfo {year} {2025})}\BibitemShut {NoStop}%
\bibitem [{\citenamefont {Amick}\ \emph {et~al.}(2005)\citenamefont {Amick}, \citenamefont {Gendreau}, \citenamefont {Busch},\ and\ \citenamefont {Gordon}}]{amick2005evolving}%
  \BibitemOpen
  \bibfield  {author} {\bibinfo {author} {\bibfnamefont {H.}~\bibnamefont {Amick}}, \bibinfo {author} {\bibfnamefont {M.}~\bibnamefont {Gendreau}}, \bibinfo {author} {\bibfnamefont {T.}~\bibnamefont {Busch}},\ and\ \bibinfo {author} {\bibfnamefont {C.}~\bibnamefont {Gordon}},\ }in\ \href@noop {} {\emph {\bibinfo {booktitle} {Buildings for Nanoscale Research and Beyond}}},\ Vol.\ \bibinfo {volume} {5933}\ (\bibinfo {organization} {SPIE},\ \bibinfo {year} {2005})\ pp.\ \bibinfo {pages} {16--28}\BibitemShut {NoStop}%
\end{thebibliography}%

\end{document}